\documentclass[runningheads]{llncs}

\usepackage[T1]{fontenc}
\usepackage{graphicx}

\usepackage{cite}
\usepackage{algorithmic}
\usepackage{graphicx}
\usepackage{textcomp}
\usepackage{xcolor}
\usepackage{comment}
\usepackage{tikz}
\usepackage{aeguill}
\usepackage[frozencache,cachedir=.]{minted}
\usepackage{enumitem}
\usepackage{parskip}
\usepackage{algorithm}
\usepackage{xcolor}
\usepackage{float}
\usepackage{subcaption}
\usepackage{hyperref}
\usepackage{caption}
\usepackage{color}

\begin{document}
\title{Methodology for Test Case Allocation based on a Formalized ODD}
%
\author{Martin Skoglund\inst{1}\orcidID{[0000-0001-6901-4986} \and
Fredrik Warg\inst{1}\orcidID{0000-0003-4069-6252} \and Anders Thorsén\inst{1}\orcidID{0000-0001-7933-3729} \and
Sasikumar Punnekkat\inst{2}\orcidID{0000-0001-5269-3900} \and Hans Hansson\inst{2}\orcidID{0000-0002-7235-6888}}
\authorrunning{M. Skoglund et al.}
%
\institute{Department of Electronics, RISE Research Institutes of Sweden, Borås, Sweden
\email{martin.skoglund@ri.se,fredrik.warg@ri.se,anders.thorsen@ri.se} \\
\and
MRTC, Mälardalen University, Västerås, Sweden
\email{sasikumar.punnekkat@mdu.se,hans.hansson@mdu.se} \\
}
\maketitle              
\begin{abstract}
The emergence of Connected, Cooperative, and Automated Mobility (CCAM) systems has significantly transformed the safety assessment landscape. Because they integrate automated vehicle functions beyond those managed by a human driver, new methods are required to evaluate their safety. Approaches that compile evidence from multiple test environments have been proposed for type-approval and similar evaluations, emphasizing scenario coverage within the system’s Operational Design Domain (ODD). However, aligning diverse test environment requirements with distinct testing capabilities remains challenging. \\
This paper presents a method for evaluating the suitability of test case allocation to various test environments by drawing on and extending an existing ODD formalization with key testing attributes. The resulting construct integrates ODD parameters and additional test attributes to capture a given test environment’s relevant capabilities. This approach supports automatic suitability evaluation and is demonstrated through a case study on an automated reversing truck function. The system's implementation fidelity is tied to ODD parameters, facilitating automated test case allocation based on each environment’s capacity for object-detection sensor assessment.
\end{abstract}

\section{Introduction} \label{sec:intro}
The safety assurance of Connected, Cooperative, and Automated Mobility (CCAM) systems is a critical challenge for their widespread adoption. As higher levels of automation are pursued, traditional validation through real-world testing becomes impractical due to the immense number of scenarios required. In the automotive field, this is commonly called the "billion-miles" challenge~\cite{kalraDrivingSafetyHow2016a} but extends to any domain with automation ambitions. An appropriate mix of physical and virtual testing has emerged as a more feasible solution in such contexts. A blended physical and virtual strategy is, therefore, the practical alternative. 

Despite these efforts, a significant gap remains between high-level schematic descriptions and practical guidance in concrete methods. The lack of a practical validation hampers the safe and large-scale deployment of CCAM technologies, with many still under development or recently introduced. 

Today, scenario-based testing for automated driving is growing in importance and prevalence. However, it is still a challenge to determine if a test suite sufficiently covers the ODD~\cite{weissensteinerODD2023}. Part of solving this is to develop systematic methods to align scenario requirements with distinct test environment capabilities. Integrating ODD parameters with test attributes can address this gap by enabling automated test case allocation to appropriate test environments. A subcategory of this topic is an external assessment of the appropriateness of such allocations, as required by functional safety standards~\cite{iso26262}. This assessment is similarly complex for the reasons that hamper the initial allocation, particularly scope, and link to intended context and test environment appropriateness \cite{templates}.

\begin{figure*}[htbp!]
    \begin{minipage}{1\textwidth}
        \centering
        \includegraphics[width=1\textwidth]{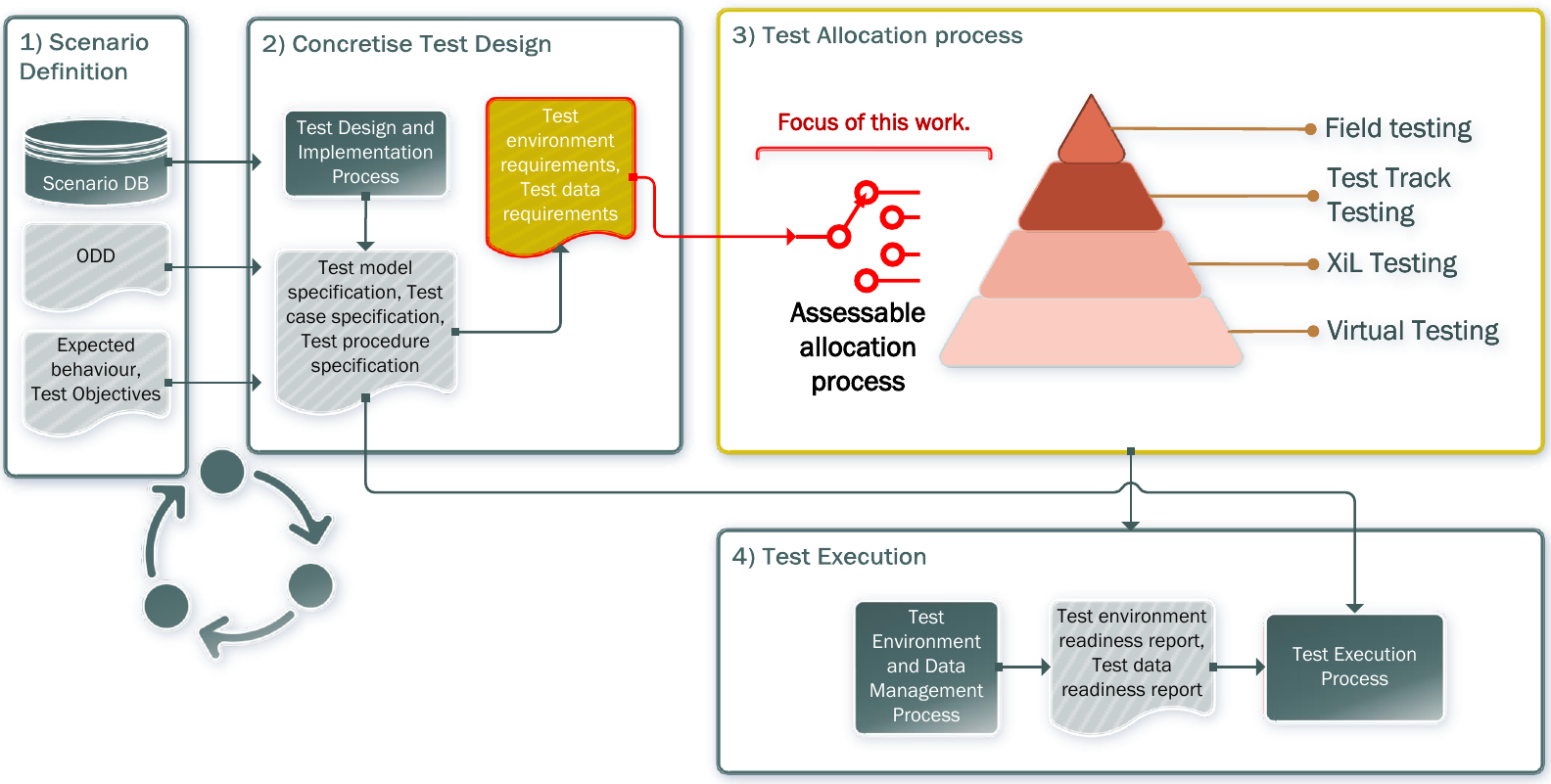}
        \caption{Schematic overview of a scenario‑based safety‑assurance process highlighting the allocation step.}
        \label{fig:SAF}
    \end{minipage}\hfill
\end{figure*}

Building on Road vehicles - Test scenarios for automated driving systems ISO 3450X~\cite{iso34501,iso34502,iso34503} and an ODD formalization~\cite{ODD_PKL_2025}, this paper proposes an automated test case allocation process centered on extending an ODD object with test environment attributes. The ODD parameters alone do not fully reflect a test environment’s capacity to address hazards, complexity, and fidelity. Accordingly, we extend the ODD concept with additional test environment attributes, forming a unified structure that better aligns test scenario requirements with test environment capabilities.

The approach confines test requirements to specified capabilities, allowing for the evaluation and automated allocation of test cases based on each environment’s capacity to provide relevant safety evidence. Cost and scheduling considerations, which lie outside the scope of functional safety, are excluded to maintain focus on safety-specific concerns.

This study builds on the ISO 3450x scenario framework and a recent formalization of operational design domains to propose an automated allocation method that augments the ODD with test‑environment descriptors~\cite{ODD_ext_git}. Confining test requirements to declared capabilities enables objective allocation of scenarios to those environments that can produce credible safety evidence; this claim is illustrated with an automatic reversing‑truck case study supported by an open‑source implementation. The remainder of this paper is organized as follows. 
Section \ref{sec:background} reviews related work on scenario‑based safety assurance. Section \ref{sec:odd} details the proposed methodology for test allocation based on a formalized ODD. Section \ref{sec:example} presents the reversing‑truck case study, and Section \ref{sec:conclusions} summarises the principal findings and outlines avenues for future research.


\section{Background and Related Work} \label{sec:background}
Automated driving functions exemplify the broad challenges associated with CCAM safety assurance. Growing complexity and variant diversity lead to exponentially increasing testing demands that exceed the capacity of conventional requirement-based approaches. 
Many initiatives adopt a scenario-based perspective to address the increased complexity, at least for top-level testing~\cite{iso34501,iso34502,iso34503}. This practice can improve coverage of diverse and potentially unforeseen corner cases while enabling reuse across different functionalities. However, it also creates challenges in ensuring completeness. Scenario-based tests often require significant computational and organizational resources for test design, execution, and assessment across heterogeneous environments, and the principal challenge lies in implementing these methods at scale~\cite{RiedmaierSurveyScenarioBasedSafety}. The approach offers increased flexibility in adapting to evolving test requirements by decoupling scenarios from complex, difficult-to-maintain test code. Its practical relevance is underscored by UNECE Regulation No. 157, which governs automated lane-keeping systems and highlights the importance of scenario-based testing in ensuring robust system performance~\cite{No157}.
Scenario-based safety assurance approaches can be seen as an extension of the dynamic testing described in Software and systems engineering —
Software testing ISO 29119~\cite{iso29119_2022}, which more comprehensively addresses processes, documentation, techniques, and test management in software testing, a wealth of information to be drawn upon in areas where ISO 3450x lacks details. Additionally, the structured use of high-dimensional ODD parameter data for automated testing in automated driving supports the data-driven intelligent transportation systems approach, which leverages diverse, large-scale data to enhance safety, efficiency, and decision-making~\cite{DDE}.

Fig.\ref{fig:SAF} schematically illustrates four main stages of a scenario-based safety assurance process focusing on putting the test case allocation method in context, in line with approaches such as~\cite{natm, sunriseSAF, NTHSA2018}. The first stage, Scenario Identification (Fig.\ref{fig:SAF} Stage 1), defines the ODD and the system's expected behavior. Relevant scenarios are sourced from a database and aligned with test objectives. The second stage, Concrete Test Design (Fig.\ref{fig:SAF} Stage 2), involves translating these high-level scenarios into detailed test cases and specifying the necessary test environment and data requirements, following guidance from ISO 29119~\cite{iso29119_2022}. A test specification encompasses all test design elements, including the test cases, procedures, and requisite environments. \\
The third stage, the Test Case Allocation Process, addresses the growing need to manage large, parameterized test suites and integrate evidence from multiple test environments. Test environments are generally categorized as field testing, test track testing, XiL testing, or fully virtual testing, each having different attributes.

The allocation aims for effectiveness—ensuring that tests produce credible safety evidence—and efficiency—matching scenarios to environments suited to the required capabilities. Readiness reports (as described in ISO 29119) record environment status, data availability, resource planning, scheduling, risk assessments, and operational constraints. As aims for a method that focuses on safety and needs to be agnostic to the technology up to point in interface with different ODD parameters, in contrast to the similar methods proposed by Striemle et al.~\cite{steimle2022}. \\
The final stage, Test Execution (Fig.\ref{fig:SAF} Stage 4), proceeds once test cases have been allocated to specific environments. It involves verifying the environment and data are ready, executing test cases, and reporting the results. As exemplified in Section \ref{sec:example}, environments should maintain validated parameter ranges, repeating tests that exceed or approach these boundaries in more reliable settings to ensure credible outcomes. Machine-readable scenarios and ODD specifications reduce errors in preparation and execution by confirming that collected data meets the requirements for evaluation and coverage.

\section{Methodology for Test case Allocation based on a Formalized ODD (METAFODD).} \label{sec:odd}
In the context of the construction of a test case allocation methodology, we leverage an ODD taxonomy construct consistent with ISO 34503~\cite{iso34503}, as well as the formalizing ODDs by the use of the Pkl~\cite{Pkl} language, as proposed by Skoglund et al.~\cite{ODD_PKL_2025}. From that work, we have a hierarchical taxonomy ODD definition, an inclusive ODD, where parameters must be explicitly specified. Our work of refining test attributes into a minimal essential set for the initial allocation process is detailed in~\cite{D33}, emphasizing the key factors required to achieve the intended evaluation objectives. \textbf{Test environment attributes}: These include several aspects related to the capacity of the testing system: 
    \begin{itemize} \label{attributes}
        \item \textbf{Safety Hazard Mitigation Capability}: The ability to minimize potential hazards, which could pose risks to participants, including safety drivers and experiment observers, commonly associated with track testing.
        \item \textbf{Test Complexity Capability}: The degree of complexity involved in testing, including the facility's ability to accommodate diverse test elements, orchestration, and ODD conditions.
        \item \textbf{Test Environment Fidelity Capability}: The accuracy with which test models replicate real-world conditions, including vehicle and road user behavior, relevant to the test coverage item, i.e., what you are testing.
        \item \textbf{System Under Test (SUT) Fidelity Capability}: A metric that assesses the abstraction between a model and its intended production implementation, considering the limitations of virtual environments or test harnesses relevant to the test coverage item.
    \end{itemize}

The ODD template is extended with four additional test environment attributes. These attributes must be specified both in the test case definition, which represents the requirements and in the test environment capabilities, which represent the provider. Both sides use the same extended template to ensure comparability for validation. Each of the four test attributes is subdivided into low, medium, and high levels, reflecting incremental capability, where higher levels include the properties of the lower levels. In PKL, this extension can be represented as an addition to the ODD, as illustrated in Fig.~\ref{alg:extendodd}. Low generally indicates minimal emphasis or significant abstraction, medium corresponds to partial coverage or moderate complexity, and high denotes thorough hazard management or near-complete fidelity.

In a typical virtual environment, safety hazard mitigation and overall throughput are often high because there is no kinetic energy, and multiple tests can run in parallel. However, environment fidelity and SUT fidelity are usually lower owing to abstracted models.
In typical XiL setups safety mitigation remains high, throughput is medium, and test complexity is moderate, although environment fidelity typically remains low and SUT fidelity is high. 
Proving ground tests usually provide a high environment and SUT fidelity because they involve real vehicles and conditions. However, safety hazard mitigation and test throughput remain low, and the practical challenges of physical testing constrain test complexity. Limited safety hazard mitigation capabilities indicate that certain high-risk tests may be infeasible and should not be conducted.
This classification scheme is acknowledged as a preliminary. With the prospect of more quantitative metrics~\cite{böde} there is an opportunity to refine these categories in future research. Nonetheless, even this coarse extension to the ODD has proved beneficial in practice, verifying the soundness of pre-existing (initial) allocations scenario coverage within the ODD.

\begin{figure}
\begin{minted}[xleftmargin=2em, mathescape, linenos, fontsize=\small]{python}
#ModuleInfo { minPklVersion = "0.25.1" }
module ODD.ODD_template.pkl

import "dyn_template.pkl"
import "env_sun_ext_template.pkl"
import "scen_template.pkl"

open class odd {
  scenery: scen_template.scenery
  environment: env_sun_ext_template.environment
  dynamic: dyn_template.dynamic_elements
}

class ext_odd extends odd {
  #   1 Low, 2 Medium , 3 High
  Safety_Hazard_Mitigation: Int (isBetween(1,3))
  Test_Complexity: Int (isBetween(1,3))
  Test_Environment_Fidelity: Int (isBetween(1,3))
  SUT_Fidelity: Int (isBetween(1,3))
}
\end{minted}
\caption{Extend the PKL formalized ISO 34503 template with four test environment attributes, specified in both test case requirements and environment capabilities for valid comparison.} 
\label{alg:extendodd}
\end{figure}
Good maintainability is achieved as the Pkl templates enable reuse by importation. An example is in Fig.\ref{alg:extendodd} where one large module is split into multiple smaller ones, templates can be repeatedly turned into concrete configurations by filling in the blanks and, when necessary, overriding defaults. One can generate static configurations in one of many standard formats to configure testing tools from this dynamic base directly. The constructed ODD templates in Pkl can be found here~\cite{ODD_ext_git}.

\section{Case Study: Reversing Truck Functionality} \label{sec:example}

\begin{figure}[htb!]
 \centering
        \includegraphics[width=0.7\textwidth]{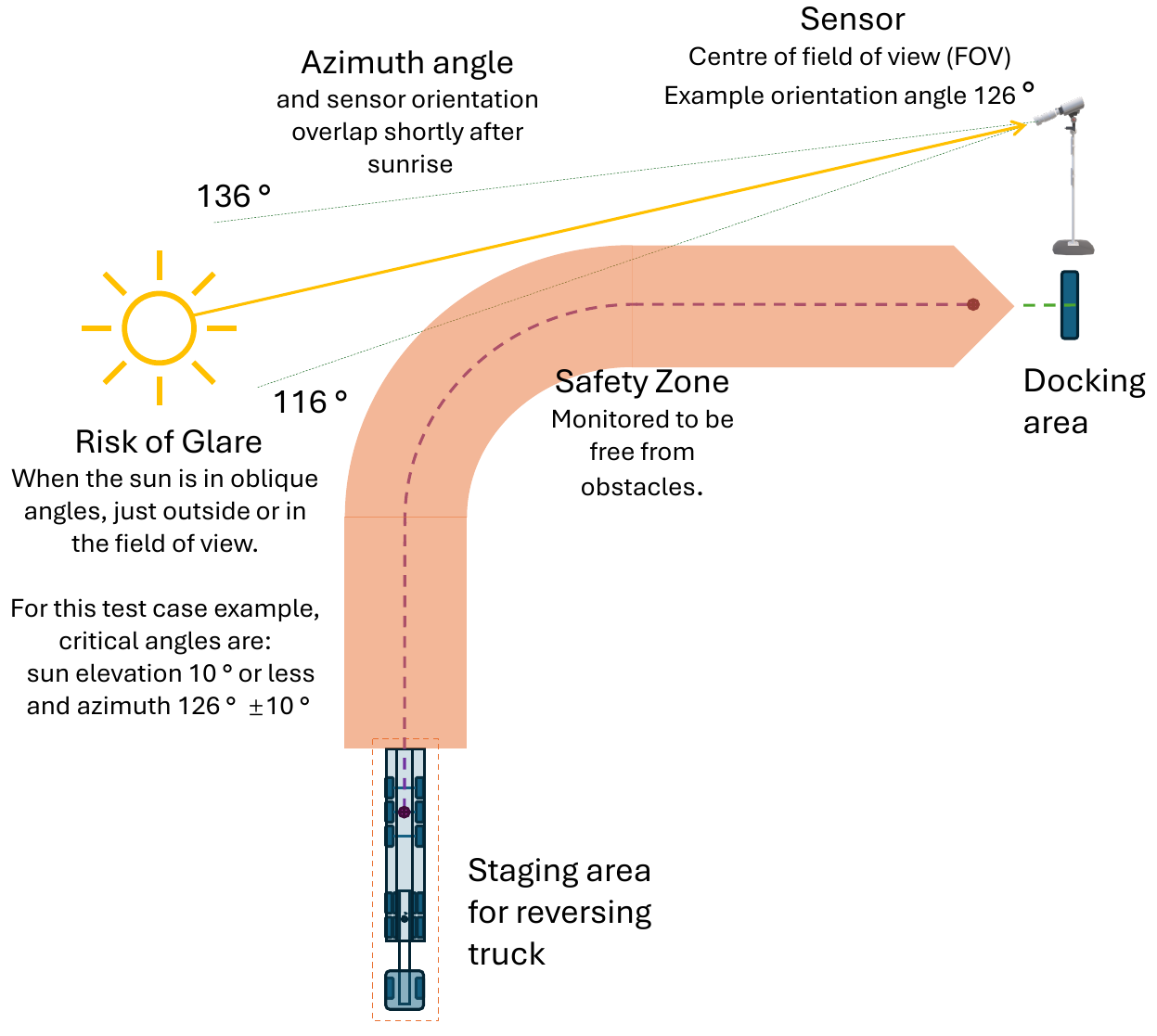}
        \caption{Test configuration for a fixed mounted camera. }
        \label{fig:glare}
\end{figure}

A case study on automated reversing of a semitrailer truck, further detailed in \cite{D72}, demonstrates how ODD parameters shape the allocation of test cases. 
Confined areas with perimeter protections and reduced unauthorized entry risks provide a well-defined operational scope to validate automated functionality in heavy vehicles. Our use case is an automated docking function of a truck to a logistic port, where the area behind the truck is monitored by a camera mounted on the hub. The camera aims to ensure the safety zone (Fig \ref{fig:glare}) is free from persons and objects. The system is defined to work during the daytime. The daytime test space and the fixed mounting of the camera will then incorporate the special problem of sun glare as defined in Fig \ref{fig:glare}, which affects object detection. 
\begin{figure*}[htbp!]
    \centering
    \begin{subfigure}[t]{0.5\textwidth}
        \centering
        \includegraphics[width=0.9\linewidth]{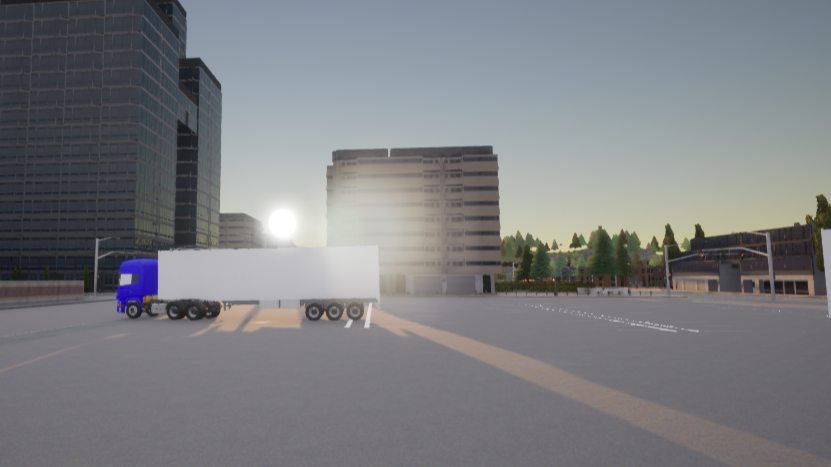}
        \caption{}
        \label{fig:carla}
    \end{subfigure}%
    \hfill
    \begin{subfigure}[t]{0.5\textwidth}
        \centering
        \includegraphics[width=0.9\linewidth]{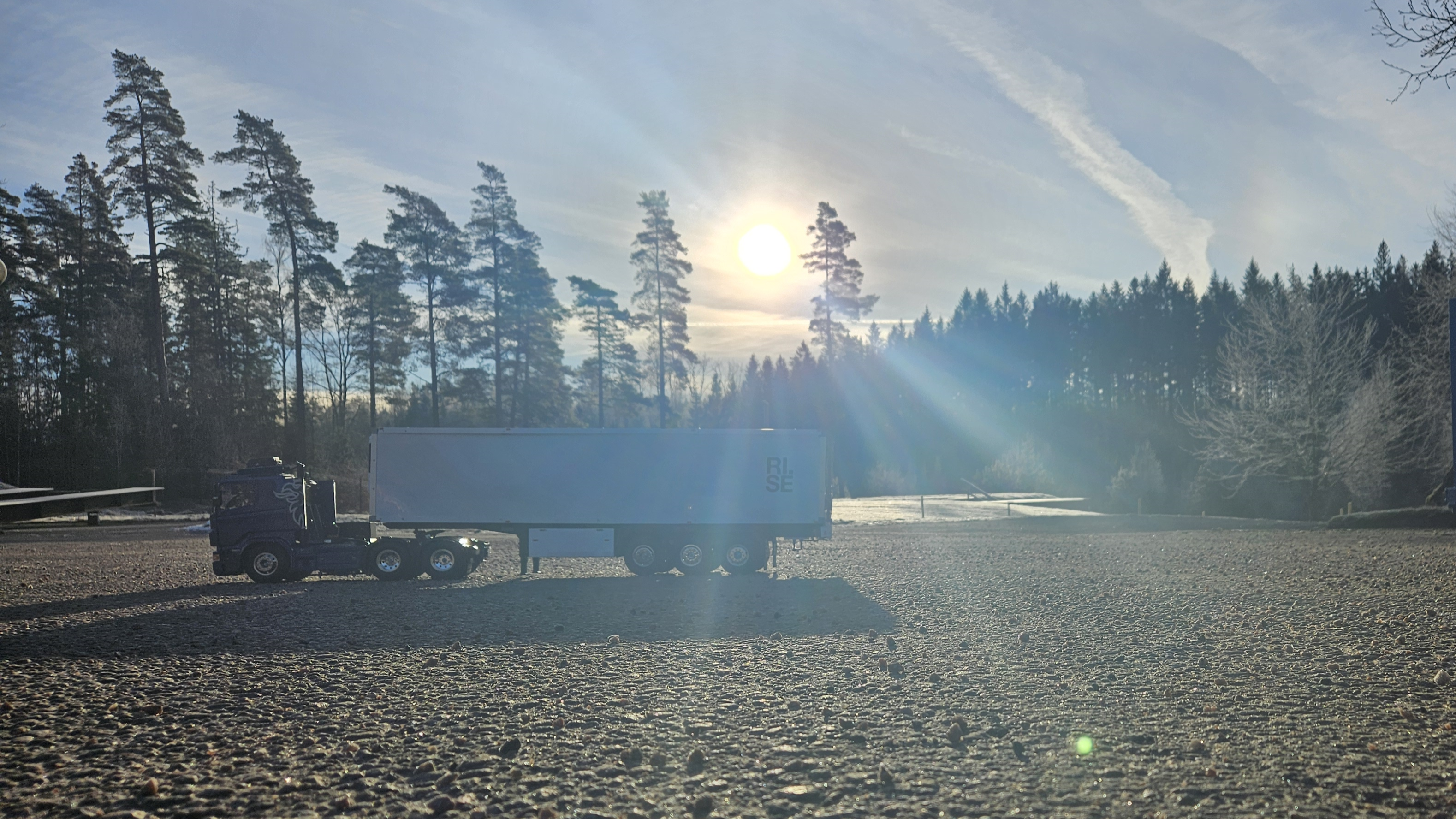}
        \caption{}
        \label{fig:scale}
    \end{subfigure}
    \caption{Illustration of low sun glare test scenarios: the simulation environment shown in (\subref{fig:carla}) uses CARLA with a sun elevation of 6°, while the hardware-in-the-loop scale truck setup in (\subref{fig:scale}) has a sun elevation of 9°.}
    \label{fig:sunglare}
\end{figure*}

\begin{figure}
    \centering
        \includegraphics[width=1\textwidth]{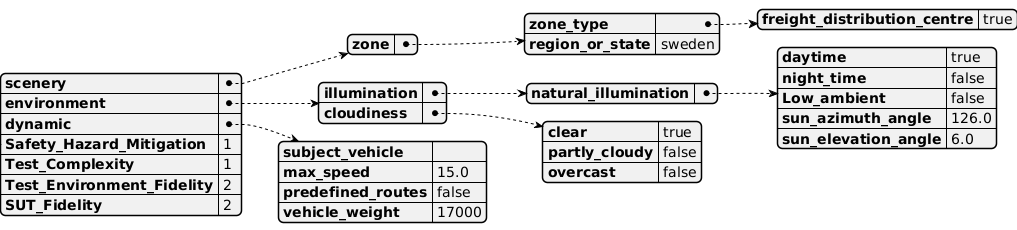} 
        \caption{A subset of the test environment requirement parameters, out of 300 ODD configurable elements. }
        \label{fig:req}
\end{figure}

In many cases, oblique angles just outside or near the field of view are the most prone to inducing reflections or scatter that manifest as glare \cite{ray}. Glare can occur over various angles depending on lens design, coatings, and light source intensity, and it must be tested in a high-fidelity environment; in this situation, a simulated environment cannot produce reliable results (see Fig. \ref{fig:carla} compared to Fig. \ref{fig:scale}), so a hardware-in-the-loop (HiL) environment will be employed. Here, camera orientation, combined with the sun's azimuth and orientation angles, defines a field of view that forms a test subspace. This expansion of the ODD to include the sun position is reflected in the ODD template and, therefore, in the test environment requirements, as shown in Fig. \ref{fig:req}. \\

Fig. \ref{fig:req} uses the template in Fig.\ref{alg:extendodd}, configured as a test environment requirement, exported to YAML format, and visualized in PlantUML \cite{plantuml}, which allows human reviewers to verify the requirements easily.

\subsection{Test environment capability, CARLA simulator}
The CARLA simulator \cite{teamCARLA} is an open-source platform used in automated driving research, offering flexibility in modeling a variety of road, weather, and lighting conditions. Fig.\ref{alg:carla} shows a subset of configurable weather parameters in CARLA, including sun azimuth and elevation angles. The test environment capabilities do include these parameters. Still, correctly modeling the SUT is essential to provide a reliable object detection test result, particularly when evaluating glare effects at oblique angles. Fully replicating complex glare conditions in simulation can be challenging, so a HiL environment will be employed where glare might be an issue to ensure reliable results.

\begin{figure}
\begin{minted}[xleftmargin=2em, mathescape, linenos, fontsize=\small]{python}
#include <WeatherParameters.h> 
WeatherParameters (
...
float in_cloudiness
float in_sun_azimuth_angle,
float in_sun_altitude_angle,  
#Same as sun_elevation_angle in ODD definition
...)
\end{minted}
\caption{Excerpt of weather parameters, it is available in CARLA.} 
\label{alg:carla}
\end{figure}

To capture the glare caveat for oblique angles, the Fig.\ref{alg:oddcapcarla}. extends SUT\_Fidelity with the conditional expression to incorporate a check on sun\_azimuth\_angle, ensuring the value lies within 126.0° ± 10.0°. It sets SUT\_Fidelity to 1 when both sun\_elevation\_angle are less than or equal to 10.0°, and sun\_azimuth\_angle remains in the given range and otherwise sets it to 2.
\begin{figure}
\begin{minted}[xleftmargin=2em, mathescape, linenos, fontsize=\small]{python}
import   "ODD_template.pkl"

odd_cap_carla: ODD_template.ext_odd = new {
  scenery {
    zone {
      region_or_state = "sweden"
      zone_type {
        freight_distribution_centre = true
      }
    }
  }
  environment {
    illumination {
      natural_illumination {
        #  Max capability
        sun_azimuth_angle = 360.0
        sun_elevation_angle = 90.0
      }
    }
  }
  Safety_Hazard_Mitigation = 3
  Test_Complexity = 3
  Test_Environment_Fidelity = 2
  #  Glare caveat for oblique angles
  #  When the risk of glare SUT_Fidelity = low 
  SUT_Fidelity = (if (
    (odd_req.environment.illumination.natural_illumination
    .sun_azimuth_angle >= 116.0)
    && (odd_req.environment.illumination.natural_illumination
    .sun_azimuth_angle <= 136.0)
    && (odd_req.environment.illumination.natural_illumination
    .sun_elevation_angle <= 10.0)
) 1 else 2)
}
\end{minted}
\caption{The CARLA test environment capability.}
\label{alg:oddcapcarla}
\end{figure}
An ODD is a complex class with a deep, non-static structure, allowing for the amendment and extension of its leaves. These leaves can take various forms, including booleans, strings, durations, data sizes, floats, and integers.

\begin{figure}
\begin{minted}[xleftmargin=2em, mathescape, linenos, fontsize=\small]{python}
# The ext_ODD_test contains test requirements and test capabilities.
# Also a method of generic comparisons that evaluate those conditions.
...
Within_CARLAs_Capabilities = genCompare.apply(odd_cap_carla, odd_req)
Within_Scaletruck_Capabilities = genCompare.apply(odd_cap_scale, odd_req)
...
C:\pkl\ODD_allocate> ./pkl eval .\ext_ODD_test.pkl
...
#  Result 
Within_CARLAs_Capabilites = false
Within_Scaletruck_Capabilites = true
\end{minted}
\caption{Automatic allocation evaluation.} 
\label{alg:compare}
\end{figure}

Because the ODD structure is large and complex, verifying whether one configured ODD is contained within another—such as comparing specific test requirements in Fig. \ref{fig:req} with the CARLA test environment capabilities in \ref{alg:oddcapcarla}—necessitates tool support. A validation method called \textit{genericCompare} is defined using the reflection property of Pkl \ref{alg:compare} \cite{ODD_ext_git}. Reflection enables querying a program's metadata, such as the classes within an assembly and the methods, fields, and properties they contain. By leveraging this capability, an intelligent recursive loop can be constructed to perform a detailed, piecewise comparison of all leaves. This method ensures that string and boolean values are checked for equality while integers and floats are compared using an equality or "less than" condition.

The proposed method for comparing two configured ODDs has several limitations. One significant limitation is handling extremes such as temperature at both ends of numeric ranges, which needs to be addressed.  Simply checking for equality or "less than" conditions may not capture the nuances of overlapping ranges or boundary conditions, limitations inherited from the specification, and also best addressed at that level. Limitations aside, the method works and can be used for both automation and assessment of allocations. 

Test criteria outlined in Section \ref{sec:odd} can be integrated with the template in Fig.\ref{alg:extendodd}. and, in conjunction with the genericCompre function (Fig.\ref{alg:compare}.), enable the comparison of test requirements (Fig. \ref{fig:req}) with environment capabilities, such as those in Fig.\ref{alg:oddcapcarla}. These elements, when integrated, form a prototype methodology for automatically allocating test cases to suitable environments.


\section{Conclusions} \label{sec:conclusions}
In conclusion, any ODD definition formalized using the Pkl language method~\cite{ODD_PKL_2025} can be extended with test environment attributes to capture test environment capabilities better. This enables automated, flexible, and scalable test allocation.

This framework-agnostic approach aligns with multi-pillar validation strategies such as NATM~\cite{natm} and SUNRISE~\cite{sunrise}, making it compatible with assurance cases that rely on heterogeneous evidence from diverse test environments. The representation permits verification of whether one ODD is subsumed by another, demonstrating its scalability and efficiency in handling extensive ODDs and ability to handle scenarios requiring finer-grained environment attributes. 

We propose and provide~\cite{ODD_ext_git} an approach that extends the ODD~\cite{iso34503} formalization in the Pkl configuration language by incorporating test environment attributes and tools for automated test case allocation, facilitating systematic and data-driven matching of scenario requirements to environment capabilities. Although still a proof of concept, this approach establishes a foundation for further refinement and broader adoption through community collaboration. Its continued development may benefit developers, assessors, tool vendors, and standardization bodies, and has the potential for wider use if its value is recognized by the research community.

Future work will examine domain-specific ODD definitions—such as those in forestry—and expand the formalization to generate test spaces that facilitate automated allocation. Efforts will also include investigating compatibility with OpenODD~\cite{OpenODD} to ensure alignment with emerging ASAM standards and explore potential integration opportunities.


\begin{credits}
\subsubsection{\ackname} We acknowledge the support of the Swedish Knowledge Foundation via the industrial doctoral school RELIANT, grant nr: 20220130. This research was carried out within the SUNRISE project and is funded by the European Union's Horizon Europe Research and Innovation Actions under grant agreement No.101069573. However, views and opinions expressed are those of the author(s) only and do not necessarily reflect those of the European Union or the European Union's Horizon Europe Research and Innovation Actions.

\subsubsection{\discintname}
The authors have no competing interests to declare relevant to this article's content. 
\end{credits}
%
%
%
\bibliographystyle{splncs04}
\bibliography{./ref/SUNRISEInitalAllocation.bib}

\end{document}